\documentclass[a4paper,11pt]{article}
\pdfoutput=1 

\usepackage{jinstpub} 

\usepackage{lineno}

\usepackage{float}
\usepackage{caption}
\usepackage{subcaption}

\title{Design of an Internet of Things - based multi-channel temperature monitoring system}

\author{B. Clerbaux,}
\author{\underline{D. Gómez de Gracia},}
\author{P. Petitjean,}
\author{Y. Yang}

\affiliation{Inter-University Institute For High Energies (IIHE VUB-ULB), Université Libre de Bruxelles, Brussels, Belgium}

\emailAdd{dgomezde83@gmail.com}

\abstract{

In the scope of the Jiangmen Underground Neutrino Observatory (JUNO) project, 6 back-end card (BEC) mezzanines connected to one BEC base board are in charge of compensating the attenuated incoming data from 48 front-end channels over 48 100-meter-long ethernet cables. Each of the mezzanines has 16 equalizers that may be subject to overheating. It is important therefore to monitor their temperature in real time. However, collecting data from a relatively large (1080) number of mezzanines is not a trivial task. In this work we propose a solution based on Wi-Fi mesh. Both the technical details and the first test results performed are be reported.

}

\keywords{Back-end card, ESP32, Wi-Fi mesh}

\proceeding{
TWEPP 2021 Topical Workshop on Electronics for Particle Physics
}

\begin{document}
\maketitle
\flushbottom

\section{Introduction}
\label{sec:intro}
\begin{figure}[h]
\begin{center}
  \includegraphics[width=0.5\textwidth]{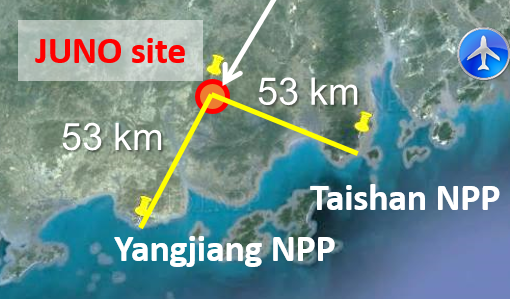}
  \caption{Location of JUNO and nuclear reactors in south of China.}
  \label{fig:JUNOsite} 
  \end{center}
\end{figure}

The Jiangmen Underground Neutrino Observatory (JUNO \cite{a}) is a multipurpose neutrino detector. Its main goal is to determine the neutrino mass hierarchy and to measure some of the neutrino oscillation parameters with unprecedented precision, using as a source electronic anti-neutrinos coming from the Yangjiang and Taishan Nuclear Power Plants (see figure \ref{fig:JUNOsite}). JUNO is located at 700 m underground and its central detector consists of 20 ktons of liquid scintillator contained in a 35 m diameter acrylic sphere that will capture neutrino interactions from different sources. Among the numerous electronic components of the detector, a back-end card (BEC) is used as a concentrator to collect trigger requests. Each of the incoming trigger request signals passes through an equalizer placed on a Mezzanine, mounted on a BEC to compensate for the attenuation due to the long cables. There are 180 BECs in the system, and a total of 17280 equalizers. Measuring the temperatures of the multiple equalizers placed on all the Mezzanines is necessary to make sure that they perform stably and don’t overheat.

\section{JUNO Electronics}
\label{sec:JUNOElectronics}
\begin{figure}[H]
  \centering
  \begin{subfigure}{0.49\textwidth}
    \includegraphics[width=1\textwidth]{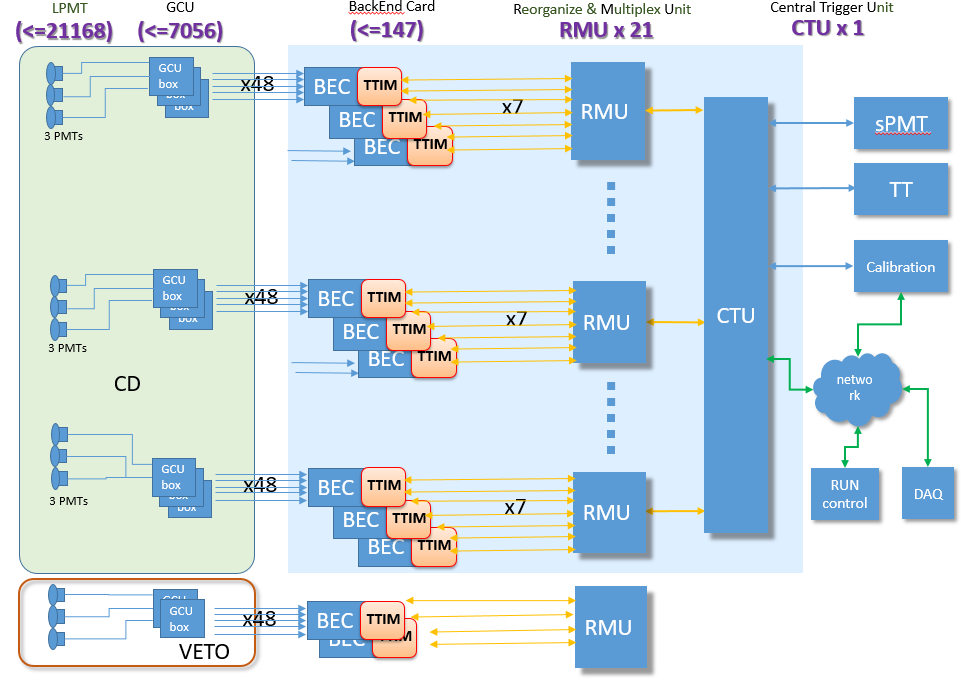}
    \caption{Schematic view of JUNO electronic system.}
    \label{fig:JUNOelectronics}
  \end{subfigure}
  \begin{subfigure}{0.49\textwidth}
    \includegraphics[width=1\textwidth]{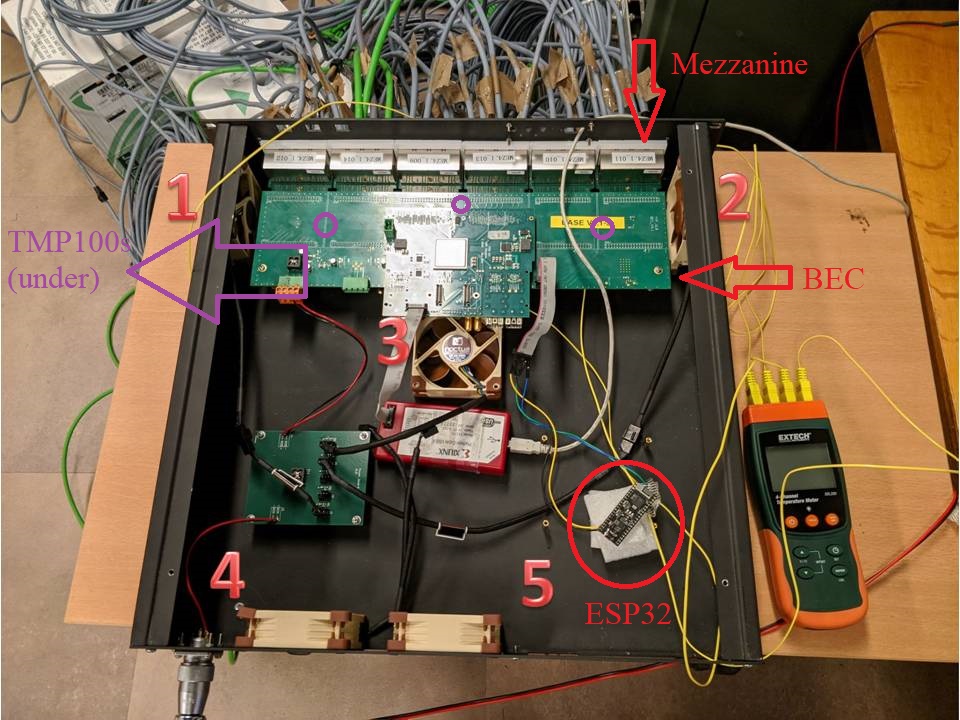}
    \caption{Open box with the 6 Mezzanines, the BEC, the TTIM mounted on the BEC, and the 5 fans (with their corresponding numbers). Main air flow enters on fan 1 and exits on fan 2.}
    \label{fig:boxinside}
  \end{subfigure}
  \caption{JUNO electronics: schematic view and main components used in this work.}
  \label{fig:three graphs}
\end{figure}

The JUNO electronics system has two parts: the "wet electronics" system and the "dry electronics" system. Both systems are inter-connected with 100m Ethernet cables \cite{b}. Figure \ref{fig:JUNOelectronics} shows how the components of both systems work together.

The wet electronics of JUNO performs basic tasks of data acquisition. The photomultiplier tubes (PMTs) collect (in groups of three) the light induced by the interaction of neutrinos with the liquid scintillator. This light is transformed into an electrical signal, which is usually called "PMT waveform" in the literature. This signal is digitized inside the global control unit (GCU), and then sent to the BEC through the synchronous link. On the other hand, the dry electronics system controls the trigger requests. The central trigger units (CTUs) compute signals to trigger data acquisition (called trigger decisions). The BECs transmit the incoming trigger decisions, which arrive through the Trigger Timing Interface Mezzanine (TTIM), towards the GCUs \cite{c}. A box is used to gather some of the elements of the dry electronics system, like the Mezzanines, the BEC and the TTIM. The exact placement of the different components is shown in figure \ref{fig:boxinside}.

\section{Engineering Analysis and Development}
\begin{figure}[h]
  \centering
  \begin{subfigure}{0.49\textwidth}
    \includegraphics[width=1\textwidth,height=3.28cm]{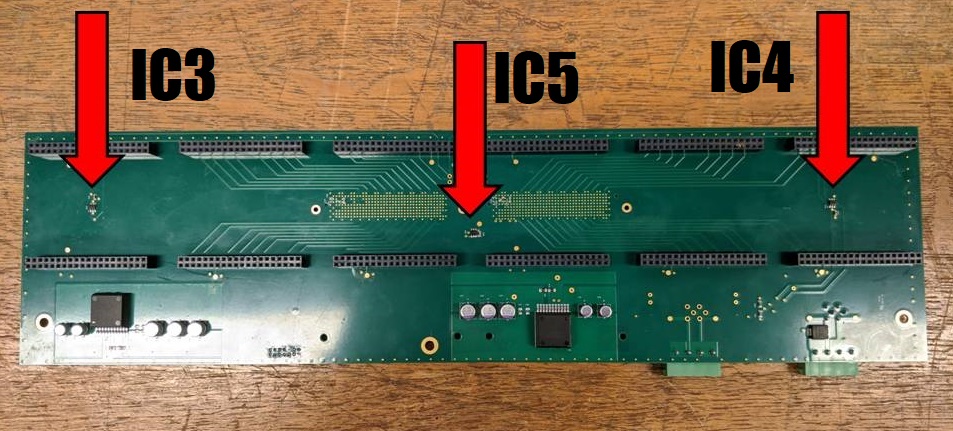}
    \caption{One BEC with three TMP100 sensors (named IC3 IC4 and IC5).}
    \label{fig:becfront}
  \end{subfigure}
  \begin{subfigure}{0.49\textwidth}
    \includegraphics[width=1\textwidth,height=3.3cm]{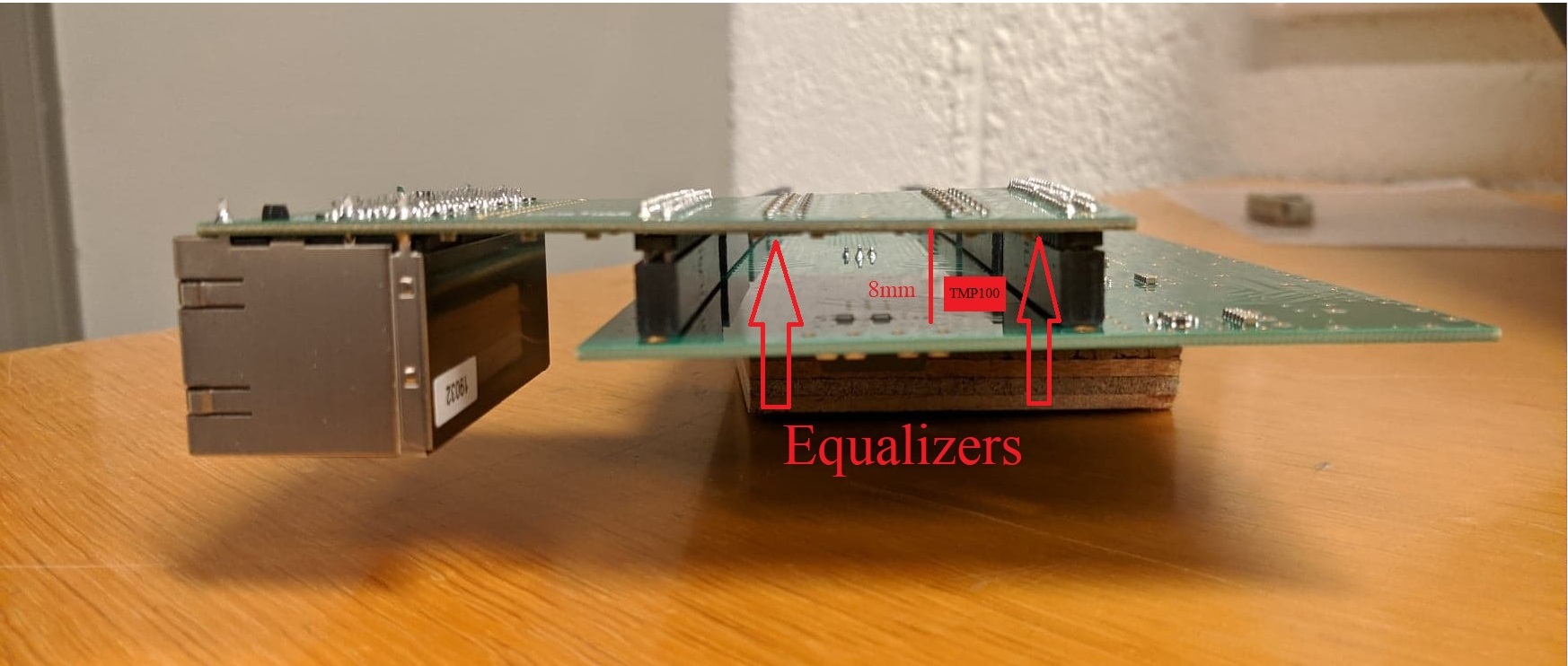}
    \caption{One BEC (below) and one mezzanine (above), 8 mm apart.}
    \label{fig:becandmezz}
  \end{subfigure}
  \caption{Two views on the BEC.}
  \label{fig:three graphs}
\end{figure}

The main goal of this work is to design a system to monitor the temperature of the 17280 equalizers present in the 180 BECs from a remote computer. Three TMP100 temperature sensors are placed on the BEC (see figure \ref{fig:becfront}). These are digital temperature sensors \cite{d} offering an average accuracy of $\pm 1$°C, a maximum resolution of 0.0625°C, and a maximum error of  $\pm 2$°C with respect to the actual temperature of the BEC's silicon on the spot the sensor is placed at. The equalizers that need to be monitored lie on the mezzanines, which are 8 mm apart from the BEC (see figure \ref{fig:becandmezz}). This means that the temperature that will be measured by the sensors is the one of the hot air above the equalizers.

\section{Implementation}
An ESP32 pico board is placed on each BEC, and it is connected to the BEC's inter-integrated circuit ($I^2C$) bus as a master device. ESP32s are development boards compatible with $I^2C$ and with Wi-Fi integration. They can communicate with the three sensors, which are connected as $I^2C$ slave devices, and read their registers at the desired rate.
Each ESP32 board should transmit its values at a frequency of 0.2 Hz, i.e. one message every 5 seconds, to the central computer in the form of messages that clearly separate the three temperature values, as well as the ID of the board where the temperatures were measured. Once they arrive to the central computer, they should be logged into a database with the time and date they were received at. The status of all the motherboards should be accessible both in real time and at any point in the monitoring timeline.

\begin{figure}[h]
  \centering
  \begin{subfigure}{0.49\textwidth}
    \includegraphics[width=1\textwidth]{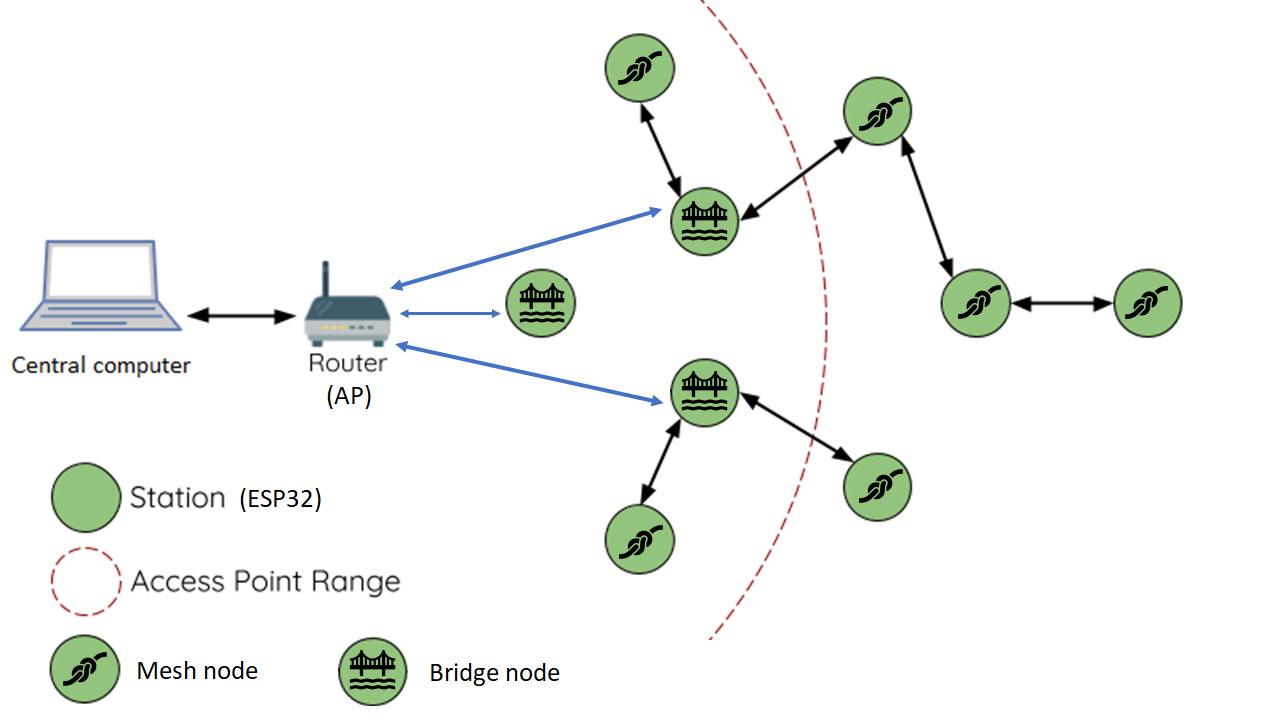}
    \caption{Wi-Fi mesh network with central computer, an arbitrary number of "Bridge" nodes, and "Mesh" nodes.}
    \label{fig:meshConnection}
  \end{subfigure}
  \begin{subfigure}{0.49\textwidth}
    \includegraphics[width=1\textwidth]{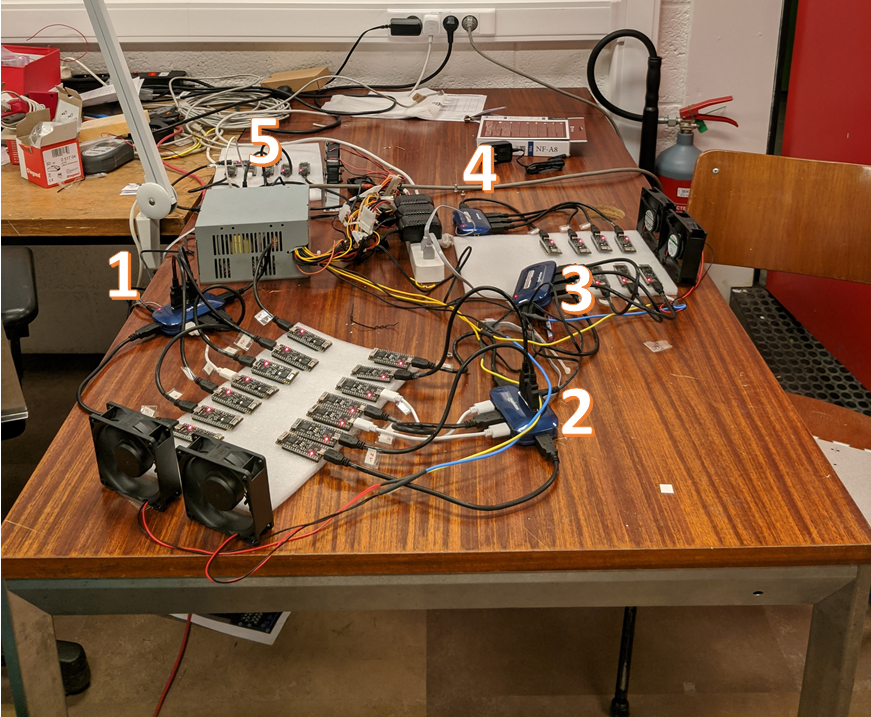}
    \caption{Reduced-scale test in the lab with 30 boards and 5 hubs (with their corresponding numbers). Transmission rate: 0.2 Hz. Reception rate: 99.06\%.}
    \label{fig:reducedScaleTest}
  \end{subfigure}
  \caption{Wi-Fi mesh network architecture - schematic view and lab installation.}
  \label{fig:three graphs}
\end{figure}

In order to achieve this, a Wi-Fi mesh network was designed (see figure \ref{fig:meshConnection}). This architecture features two kind of nodes that can connect to a single device, and accept at most 10 connections from other devices (which is the maximum amount of connections the ESP32 hardware can manage): \textit{Bridge nodes} and \textit{Mesh nodes}. The former are directly connected to an access point, and the latter connect to other nodes of the network. Previous studies (such as the one described in \cite{e}) show that transmitting messages over this kind of network should be reliable and relatively fast (less than 100 ms delay). The architecture was implemented using the PainlessMesh library \cite{f}, which automatically builds and maintains the mesh network, while allowing to incorporate new boards on the go, message sending, message broadcasting, and identification of boards.

\section{Validation and Results}
Several experiments were performed to ensure the proper functioning of the mesh network, to measure its performance, and to determine the difference between the retrieved temperatures and the real temperatures of the equalizers.

A small scale test was performed in the lab with 30 ESP32s (see figure \ref{fig:reducedScaleTest}). The goal of the test was to study the reliability and performance of the mesh network by measuring the percentage of retrieved messages and the eventual issues with the mesh formation or maintenance. The result showed that for a transmission rate of 0.2 Hz the reception rate was of 99.06\% (which means that less than 1\% of the sent messages were lost) over a period of 11 days.

\begin{figure}[h]
    \centering
    \includegraphics[width=0.48\linewidth]{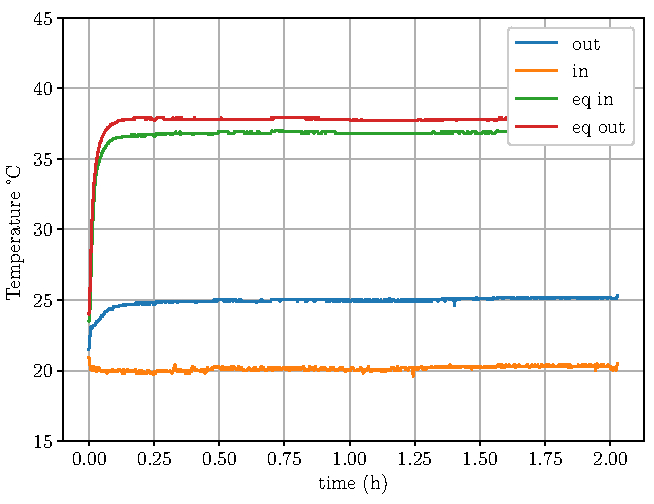}\hfill
\includegraphics[width=0.48\linewidth]{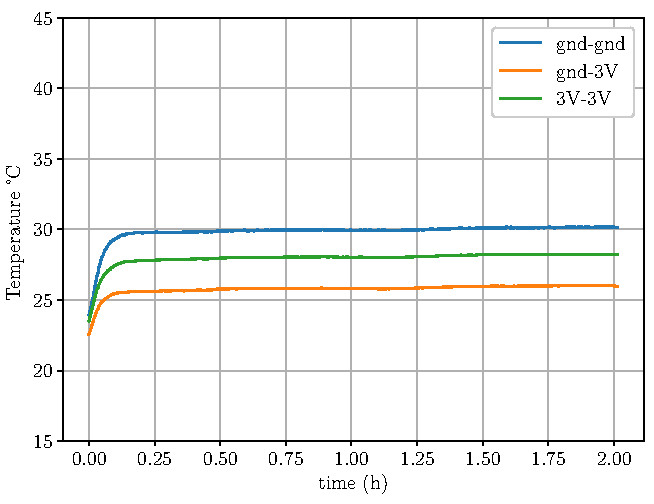}\hfill
\caption{Temperature measurement for a period of 2 h. Left: real temperatures measured with a thermocouple at fans 1 (in) and 2 (out), and at mezzanines 1 (eq in) and 6 (eq out). Right: temperatures measured by the three TMP100 sensors connected to one ESP32 (GND-GND = IC5 (center), 3V-3V = IC3 (near fan 2), GND-3V = IC4 (near fan 1)).}
\label{fig:ThermocoupleTemps}
\end{figure}

To determine the difference between the measured temperatures and the real temperatures, we placed four thermocouples on the box. Two of them on fans 1 and 2, and the two others on mezzanines 1 and 6 (which are next to fans 1 and 2, respectively, as seen in figure \ref{fig:boxinside}). These thermocouples read the temperatures of the air entering (fan 1) and exiting (fan 2) the box, and the temperatures of the equalizers on the mezzanines (\textit{eq in} and \textit{eq out}). Figure \ref{fig:ThermocoupleTemps} shows the temperatures read by the thermocouples (left) and by one ESP32 (right) during a period of 2 hours. As expected, the equalizers are 12-13°C warmer than the hot air above them that the TMP100s measure. The temperature at the fans is actually about 3-5°C lower than the temperature at the TMP100s.

\section{Conclusion}
Given the results presented in this short article, we can conclude that a Wi-Fi mesh architecture of 180 boards can be implemented if a sufficient number of Bridge nodes is chosen. This network can reliably transmit to a central computer more than 99\% of the messages each board produces if the frequency of transmission is kept at one message every five seconds or more.



\begin{thebibliography}{99}

\bibitem{a}
Jiangmen {Underground} {Neutrino} {Observatory}.
Available at: \url{http://juno.ihep.cas.cn/}

\bibitem{b}
Barbara Clerbaux and Shuang Hang and Pierre-Alexandre Petitjean and Peng Wang and Yifan Yang, \emph{Automatic test system of the back-end card for the JUNO experiment}, (2020).
Available at: \url{https://arxiv.org/abs/2011.06823}

\bibitem{c}
Yifan Yang and Barbara Clerbaux, \emph{Design of a common verification board for different back-end electronics options of the JUNO experiment}, (2018).
Available at: \url{https://arxiv.org/abs/1806.09698}

\bibitem{d}
{TMP100} data sheet, product information and support {\textbar} {TI}.com.
Available at: \url{https://www.ti.com/product/TMP100\#design-development\#\#all}

\bibitem{e}
Chia, Yoppy and Arjadi, R and Setyaningsih, Endah and Wibowo, Priyo and Sudrajat, Muhammad, \emph{Performance Evaluation of ESP8266 Mesh Networks}, (2019).
Journal of Physics: Conference Series,10.1088/1742-6596/1230/1/012023.

\bibitem{f}
A painless way to setup a mesh with ESP8266 and ESP32 devices. Available at: \url{https://gitlab.com/painlessMesh/painlessMesh/-/wikis/home}





\end{thebibliography}
\end{document}